\documentstyle[twocolumn,epsfig,aps]{revtex}

\global\firstfigfalse

\begin{document}
\draft	

\title{Two Modes of Solid State Nucleation --- Ferrites, Martensites and
Isothermal Transformation Curves}

\author{Madan Rao$^{\star}$ and Surajit Sengupta$^{\dagger}$}

\address{$^{\star}$Raman Research Institute, C.V. Raman Avenue, Bangalore
560080, India\\
$^{\dagger}$S.\ N.\ Bose National Centre for Basic Sciences, Block JD,
Sector III, Salt Lake, Calcutta 700091, India}

\date{\today}

\maketitle

\medskip 

{\bf When a crystalline solid such as iron is cooled across a structural
transition, its final microstructure depends sensitively on the cooling
rate\cite{CAHN,NISH}. For instance, an adiabatic cooling across the
transition results in an equilibrium `ferrite', while a rapid cooling gives
rise to a metastable twinned `martensite'\cite{NISH}. 
There exists no
theoretical framework to understand the dynamics and conditions under which
both these microstructures obtain.
Existing theories of martensite dynamics\cite{KACH,BALES,REID,ZYL}
describe this transformation
in terms of elastic strain, without any explanation for the occurence of the
ferrite.
Here we provide evidence for the crucial
role played by non-elastic variables, {\it viz.},
dynamically generated interfacial defects. A molecular
dynamics (MD) simulation of a model 2-dimensional (2d) solid-state
transformation reveals two distinct modes of nucleation depending on the
temperature of quench. At high temperatures, defects generated at the
nucleation front relax quickly giving rise to an isotropically growing
`ferrite'. At low temperatures, the defects relax extremely slowly, forcing
a coordinated motion of atoms along specific directions. This results in a
twinned critical nucleus which grows rapidly at speeds comparable to that of
sound. Based on our MD results, we propose a solid-state nucleation theory
involving the elastic strain and non-elastic defects\cite{PRL,CURRENT}, which
successfully describes the
transformation to both a ferrite and a martensite. Our work provides useful insights
on how to formulate a general dynamics of solid state
transformations.}

We study a 2d model system which exhibits exactly two distinct equilibrium
solid phases --- a square and a triangular phase (with a rhombic unit
cell) using an MD simulation on particles (`atoms') interacting with an
effective 2- and 3-body potential (see Fig.\ 1). By tuning the potential
parameters, we induce a square-to-triangular solid-state transformation at
constant temperature. Figure 1 shows two typical `quenches' (referred to
as high temperature $T_1$ and low temperature $T_2$ quenches) starting
from the equilibrium square phase. In both cases following the quench, the
product phase is formed by a process of nucleation and growth (Fig.\ 1
(inset)), but while the $T_1$ nucleation is homogeneous, the $T_2$
nucleation needs to be seeded by a single defect (heterogeneous).
Nucleation of a triangular region within the square phase gives rise to
vacancies/interstitials due to atomic mismatch, which relax with a
temperature dependent diffusion coefficient\cite{CAHN}. 
Note that (a) the use of effective potentials rather than `realistic' 
atomic potentials allows us to probe late times and large sizes\,; (b)
the experimentally more accurate quench schedule is a cut through the $T-v_3$
plane, however the essential physics is unaltered by our constant $T$
quench allowing for a more transparent interpretation of the MD results.

In an earlier paper\cite{ELASTIC}, we had presented a coarse-graining procedure
which starts from a microscopic description of a solid and constructs continuum
variables relevant for a description of the solid at larger length scales.
Thus we define coarse-grained dynamical {\it elastic} variables from the
instantaneous displacements ${\bf u}$ of the particles from the ideal
square lattice\cite{ELASTIC} --- nonlinear strain
$\epsilon_{ij}=1/2(\partial_iu_j+\partial_ju_i+\partial_i u_l \partial_j
u_l)$, density fluctuation about the mean $\langle \rho\rangle$, $\phi =
(\rho - \langle \rho \rangle)/\langle \rho \rangle$ (we shall refer to
this as the vacancy/interstitial field) and dislocation density (density
of 5-7 disclination pairs).

In the $T_1$ quench the vacancies/interstitials diffuse fast and
annihilate each other, thus encouraging the triangular phase to grow
isotropically (Fig.\ 2a) with a size $R$ that grows linearly with time
$t$. The critical nucleus is untwinned leading to a polycrystalline
triangular solid (`ferrite').

In the $T_2$ quench the seed vacancy does not diffuse away\,;  dominant
fluctuations correspond to a coordinated movement of a line segment of
atoms along the square axes (Fig.\ 3a).  Once these fluctuations get large
enough, they grow into a twinned critical nucleus with the twinning plane
along one of the square axes. The twinned nucleus grows anisotropically,
with a constant tip speed roughly equal to that of longitudinal sound
(Fig.\ 3b). The vacancies and interstitials while staying separated from
each other, are found concentrated at the interface of the growing nucleus
(Figs.\ 3c). In the direction transverse to the twinning `plane', the
nucleus grows by step formation (Fig.\ 3a).

The MD simulation highlights the crucial role played by defects in
determining the mode of nucleation and the nature of the critical nucleus.
From these simulations we learn that the relevant variables to describe
the dynamics of nucleation should include {\it both} the strain
$\epsilon_{ij}$\cite{KRUM} and the defect density $\phi$. We now describe
an analytic theory of solid-state nucleation for which the simplest
dimensionless free-energy functional describing the first-order square
($<\epsilon_{xy}> = 0$) to rhombus ($<\epsilon_{xy}> = \pm e_0$)
transition at a fixed degree of undercooling $a$ has the form,
\begin{equation}
F = \int_{x,y} \,(\nabla w_{ij})^2 + a \epsilon_{xy}^2 + b
\epsilon_{ii}^2 -
\epsilon_{xy}^4 + \epsilon_{xy}^6 + \gamma \,\phi^2 + C \phi
\,\epsilon_{ii}
\label{eq:landau}
\end{equation}
where $w_{ij} \equiv \partial_i u_j$ is the deformation tensor. A quench
across this transition to where the square phase is metastable, nucleates
a small `droplet' of the rhombus. Following the MD, we simultaneously
create an envelope of $\phi$ around the droplet\,; this is the initial
condition for the dynamical equations in the displacements ${\bf
u}$\cite{PRL,MPP} and $\phi$\,: $\partial_t \phi + {\bf v}\cdot \nabla
\phi = D_v \nabla^2 (2\gamma\phi+C \epsilon_{ii})$, where the microscopic
vacancy diffusion coefficient has an Arrenhius form $D_{v} = D_{\infty}
\exp (-A/k_BT)$, with an activation energy $A$\cite{CAHN}. The velocity
field ${\bf v}={\dot {\bf u}}+ \Gamma C \nabla \phi$, where $\Gamma$ is
the microscopic kinetic coefficient of ${\bf u}$.

For a high temperature quench $D_v$ is large, so $\phi$ relaxes much
faster than the time for the critical nucleus to form\,; subsequently the
nucleus grows isotropically into an equilibrium triangular solid
(`ferrite')\cite{PRL,CURRENT} with $R \sim t$.

For intermediate temperature quenches $D_v$ is smaller\,; the nature of
the critical nucleus is determined by computing the time taken for
critical nucleation $\tau_n$ (`first-passage-time'). Apriori we do not
know whether the critical nucleus is twinned or not, and so we perform a
variational calculation for both cases\,; the `true critical nucleus' is
the one for which $\tau_n$ is smaller.

In the linearised equation for $\phi$, the velocity field (ignoring
lattice vibrations) reduces to the front velocity ${\bf v}=v_{f} {\hat
{\bf n}} \,\delta({\bf r}-{\bf R}(t))$, where ${\bf R}(t)$ is the position
of the moving interface and ${\hat {\bf n}}$ is the unit outward normal to
the interface. Thus in the reference frame of the interface, the dynamics
of $\phi$ is diffusive.  Since the relaxation time to the local minima of
Eq.\ (\ref{eq:landau}) is smaller than barrier hopping times, we may treat
$\epsilon_{ij}$ as `slaved' to $\phi$ in the calculation of the barrier
height\cite{TTT}. We then use a Kramers' formula to evaluate $\tau_n$,
thus apart from unknown prefactors, $\tau_{n} = \Gamma^{-1}\,\,\exp(\Delta
E^*/k_BT)$, where $\Delta E^{*}$ is the $\phi({\bf r},t)$-dependent
critical barrier energy. Figure 4 clearly shows that for high temperature
quenches (low undercooling), the true critical nucleus is a `ferrite',
while for a low temperature quench it forms a twinned bicrystal. We
successfully reproduce two distinctive features of the well known $0$\%
isothermal transformation curves of martensites --- the horizontal
transformation curve beyond a well defined `start' temperature $M_s$ {\it
independent} of $D_{\infty}$, and a ferrite nose. It should be remarked that such
a calculation could not even be addressed within the conventional viewpoint.

What happens after the twinned critical nucleus is formed ?  It is
difficult to say based on our early time MD studies, whether the growing
nucleus will eventually add on more twins or whether it will continue as a
bicrystal with a single twin interface. On the other hand we may analyse
the late time dynamics within our analytic formalism. In the limit
$D_{v}\rightarrow 0$ when $\phi$ remains frozen at the moving interface, a
constrained variational calculation\cite{PRL,CURRENT}, shows that the
growing nucleus finds it favourable to add more twins. The free energy of
an anisotropic inclusion of length $L$ and width $W$, with $N$ twins along
the length is given by $F = {\Delta F} L W + \sigma_{pp} (L+W) +
\sigma_{tw} N W + \beta (L/N)^2 N $, where $\Delta F<0$ is the free energy
difference between the square and triangular phases, $\sigma_{pp}$ and
$\sigma_{tw}$ are the surface tensions of the square-rhombus and the twin
interfaces respectively, and the last term is the $\phi$ contribution at
the square-rhombus interface. Minimization with respect to $N$ gives $L/N
\sim W^{1/2}$, a relation that is empirically known for 2d martensites
like In-Tl\cite{NISH,KRUM}. Moreover the energy of the nucleus is smaller
when oriented along specific directions, the habit `planes'\cite{CURRENT}.

The square-rhombus interface is studded with an array of vacancies such
that $\phi \to 0$ on an average, with a separation equal to the twin size.
Thus the strain decays exponentially from the interface over a distance of
order $L/N$. This `fringing field'\cite{KRUM} arises dynamically in our
calculation, rather than from an imposed boundary condition\cite{KRUM}.
Indeed throughout our analysis we do not impose specific interfacial
conditions\cite{ZYL,KRUM}, preferring to allow the solid to choose its own
dynamical path.

The theoretical formalism just presented may be generalised to aribitrary
solid-state transformations. We have found that
in order to develop a general theory for
solid state nucleation, it is necessary to go beyond nonlinear
elasticity theory. More specifically, {\it the complete
set of relevant dynamical variables should include non-elastic degrees of
freedom} like defect fields. Such non-elastic degrees of freedom
should be relevant whenever a solid undergoes large shear deformations. 
Our work provides a framework to understand the
dynamics of solid state transformations and the conditions under which
various microstructures obtain.
The methodology of translating from microscopic simulations
to coarse-grained dynamical fields should be useful in other problems in
material science such as brittle-ductile transitions. We are currently
generalising our theory to
include the dynamics of dislocations, substituitional and interstitial
impurities and a coupling to external stress to study the dynamics of
shape-memory alloys.\\

\noindent
{\bf Acknowledgements}
We are grateful to S. Mayor and S. Sastry for a careful reading of the manuscript. We
thank the participants of the 1999 workshop on `Mathematical methods in solid mechanics
and material science'
at the Isaac Newton Institute,
Cambridge, for several illuminating discussions. MR thanks the Department of
Science and Technology, India for the award of a Swarnajayanthi fellowship.\\

\noindent
Correspondence and requests for materials should be addressed to {\small
M. R.}\,\, (e-mail: madan@rri.ernet.in)

\newpage

\begin{center}
{\bf Figure Caption}
\end{center}

Fig.\ 1\, {\bf Phase diagram in the $T-v_3$ plane} at a density $\langle
\rho\rangle = 1.05$ using an MD simulation in the $NVT$ ensemble. The
particles interact via an effective potential consisting of a 2-body
$V_2(r_{ij}) = v_2 \left(\sigma/r_{ij}\right)^{12}$ and a short range
3-body $V_3({\bf r}_i,{\bf r}_j,{\bf r}_k) = v_3\, [\sin^2
(4\theta_{ijk}) + \sin^2 (4\theta_{jki})+ \sin^2 (4\theta_{kij})]$.
Particles $i$ and $j$ are separated by a distance $r_{ij}$.
$\theta_{ijk}$ is the bond angle at $j$ between triplets $(ijk)$. $V_2$
favours a triangular lattice ground state while $V_3$ is minimised when
$\theta_{ijk} = 0, \pi/4, \pi/2$ and so favours a square lattice. The
units of length and energy are set by $\sigma$ and $v_2$ respectively,
making the unit of time $\sigma \sqrt {m/v_2}$, where $m$ is the particle
mass. The MD time step is $0.001$ corresponding roughly to a real time of
$1$\,fs. The order parameter $\langle \Omega\rangle =$$(\Omega_0\,N)^{-1}
\sum_{ijk} \langle \sin^2 (4 \theta_{ijk})\rangle$ takes values $0$($1$)
in the square(triangular) phases respectively. The phase diagram was
obtained by equilibrating a system of $N=1024$ particles at various
values of $T$ and $v_3$. Arrows marked $T_1$ and $T_2$ are constant
temperature quenches at $T_1=1$ and $T_2=0.15$. The values of $v_3$ to
which the solid is quenched is $v_3=3$ and $v_3=1.2$ respectively, which
lie in the region of the phase diagram where the square is metastable
(within the dotted line). Inset shows the time evolution of $\langle
\Omega \rangle$ for the two quenches, $T_1$ (thin line) and $T_2$ (thick
line), indicating a typical nucleation process.

Fig.\ 2~~ {\bf MD snapshot at $t=10^4$ following a $T_1$ quench:}
`Ferrite' nucleus grows isotropically as a polycrystalline triangular
region.  The colours code the local values of $\langle \Omega \rangle$
going from red in the square region to blue in the triangular region.  
Inset shows x-y trajectories for $5\times 10^3 < t < 25 \times 10^3$ of
$5$ particles chosen to lie along a row ($x$-axis) at $t=0$. At these
times the particles are part of the growing nucleus, and their
trajectories reveal significant diffusive motion. Number of unit cells of
the square in each direction is $110$. The vacancy/interstitial density
$\phi$ relaxes fast and quickly averages to $0$.

Fig.\ 3~~ {\bf MD snapshot at $t=6\times 10^3$ following a $T_2$ quench:}
(a) Twinned nucleus grows anisotropically. Colour coding as in Fig.\ 2.  
Inset shows trajectories for $10^3 < t < 16.5 \times 10^3$ of $5$
particles chosen to lie along a row ($x$-axis) at $t=0$ (movement from
bottom to top of inset). At these times the particles are part of the
growing nucleus, and their trajectories reveal highly coordinated,
`military' motion characteristic of martensites. Number of cells is the
same as above. (b) Positions of $5$ particles chosen to lie along a
twin-column ($y$-axis) as a function of time. Apart from thermal
oscillations, the particle postions change only when they encounter the
moving front at $t=t_s$. A plot of the position of the $i$th particle
$y_i$ as a function of $t_s$ gives the front-velocity $v_f=9.7$. This
compares well with the sound velocity calculated from the bulk modulus in
the square phase. (c) The vacancy/interstitial density $\phi$ integrated
over $x$ as a function of $y$ (along the twinning plane) at times
$t_1=2000$ and $t_2=5000$.  The arrow shows the direction of the
displacement field resulting in an increased density at one end of the
nucleus and a reduced density at the other. Inset shows the shear strain
$\epsilon_{xy}$ integrated over $y$ showing the twin boundary at
$t_1,t_2$.

Fig.\ 4~~{\bf First-Passage-Times versus degree of undercooling $a$} for a
twinned and untwinned critical nuclei. The calculation proceeds by
recognising that a perfect triangular solid is obtained from a perfect
square by the deformation, $R'_{i} = (\delta_{ij}+\epsilon_{ij})R_{j}$,
involving a shear $\epsilon_{xy}=\epsilon_{yx}= \epsilon$ and volume
compression $\epsilon_{xx}=\epsilon_{yy}={\epsilon}^2/2$.  We therefore
parametrize $\epsilon_{ij}$ by the single function $\epsilon({\bf r},t)$.  
This admits a variational ansatz for (1) `ferrite': $\epsilon=0$ outide a
grain of size $L$ and $\epsilon=e_0$ inside, with a smooth interpolation
of width $\xi$ and (2) `twinned nucleus': $\epsilon=0$ outide a
rectangular grain of length $L$, width $W$ and $\epsilon=\pm e_0$ inside,
connected by smooth interpolations. The diffusion equation, $\partial_t
\phi = D_v \nabla^2 (2\gamma\phi+C \epsilon_{ii})$ is solved with the
initial condition $\phi({\bf r},0) = \Delta {\bf u} \cdot {\hat {\bf n}}$
($\Delta {\bf u}$ is geometrical mismatch at the square-triangle interface
and is proportional to the length of the interface). Note that $\Delta
{\bf u} =0$ at the twin interface. The free-energy of the grain at time
$t$, $E(L, W, t\,; \{\xi_i\})$ is obtained from Eq.\ (\ref{eq:landau}),
which we minimize with respect to the widths $\{\xi_i\}$ for every $L$,
$W$ and $t$. The barrier energy $\Delta E^*$ and size $L^*, W^*$ of the
critical nucleus at every time $t$, is determined by the saddle-point. The
energy of the critical nucleus decreases with time, and so a crude but
easily calculable estimate of the first-passage-time is obtained by a self
consistent solution of $\tau_n = \Gamma^{-1}\, \exp(\Delta
E^*(\tau_n))\,$.  The curves have been calculated for $D_{\infty} =
10^{14}$, $A = 7$, $\gamma=0.2$ and $C=0$. The upper dotted line
represents the equilibrium transition temperature (zero undercooling). At
small undercooling $\phi$ relaxes fast and the critical nucleus is a
`ferrite', the first-passage-time for which goes through a minimum. At
larger undercooling, the single twin nucleates faster (bold line) than the
untwinned grain (lower dotted line). The arrow shows the degree of
undercooling $M_s$ below which the twinned nucleus forms (`martensite
start' temperature).

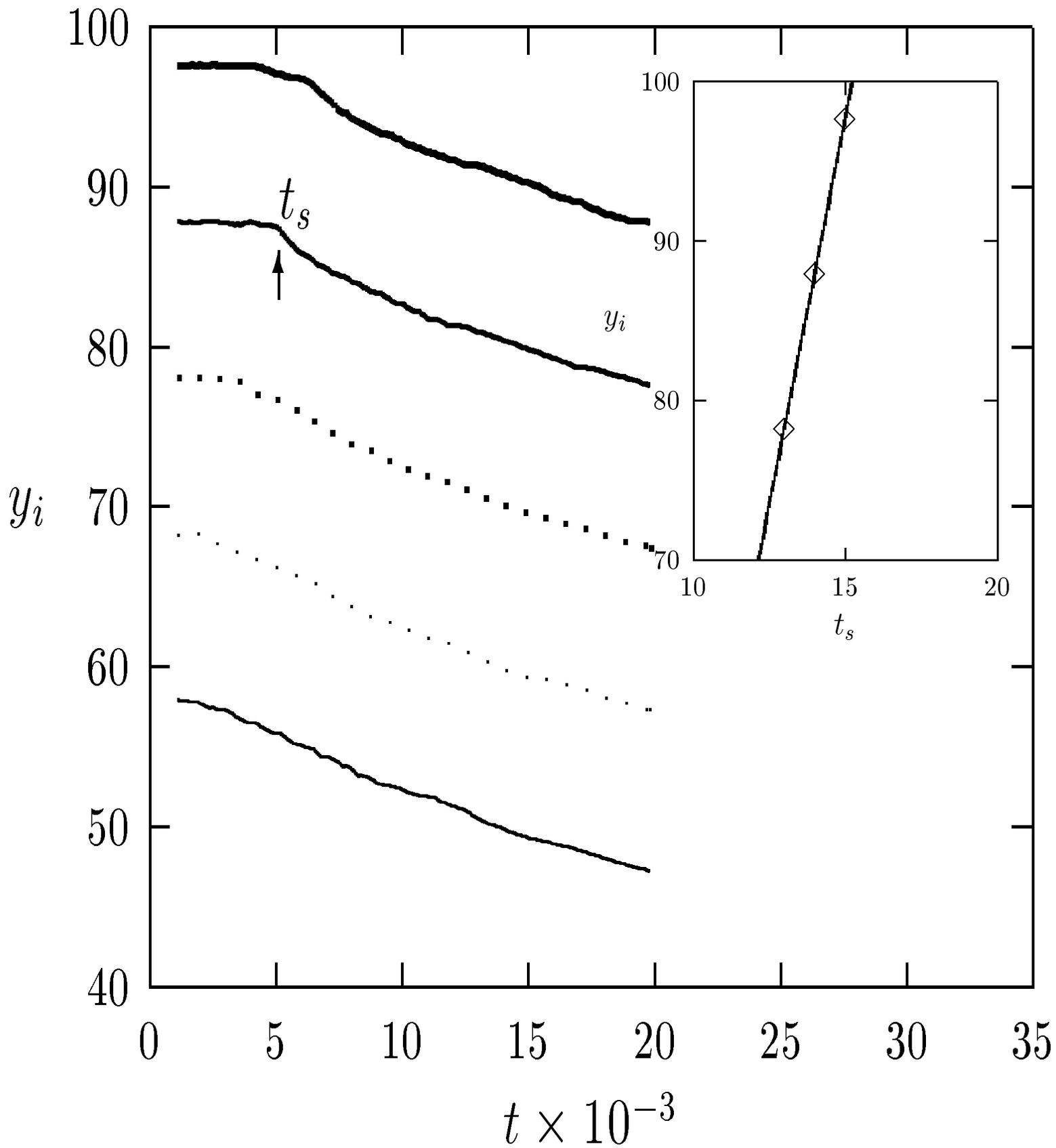
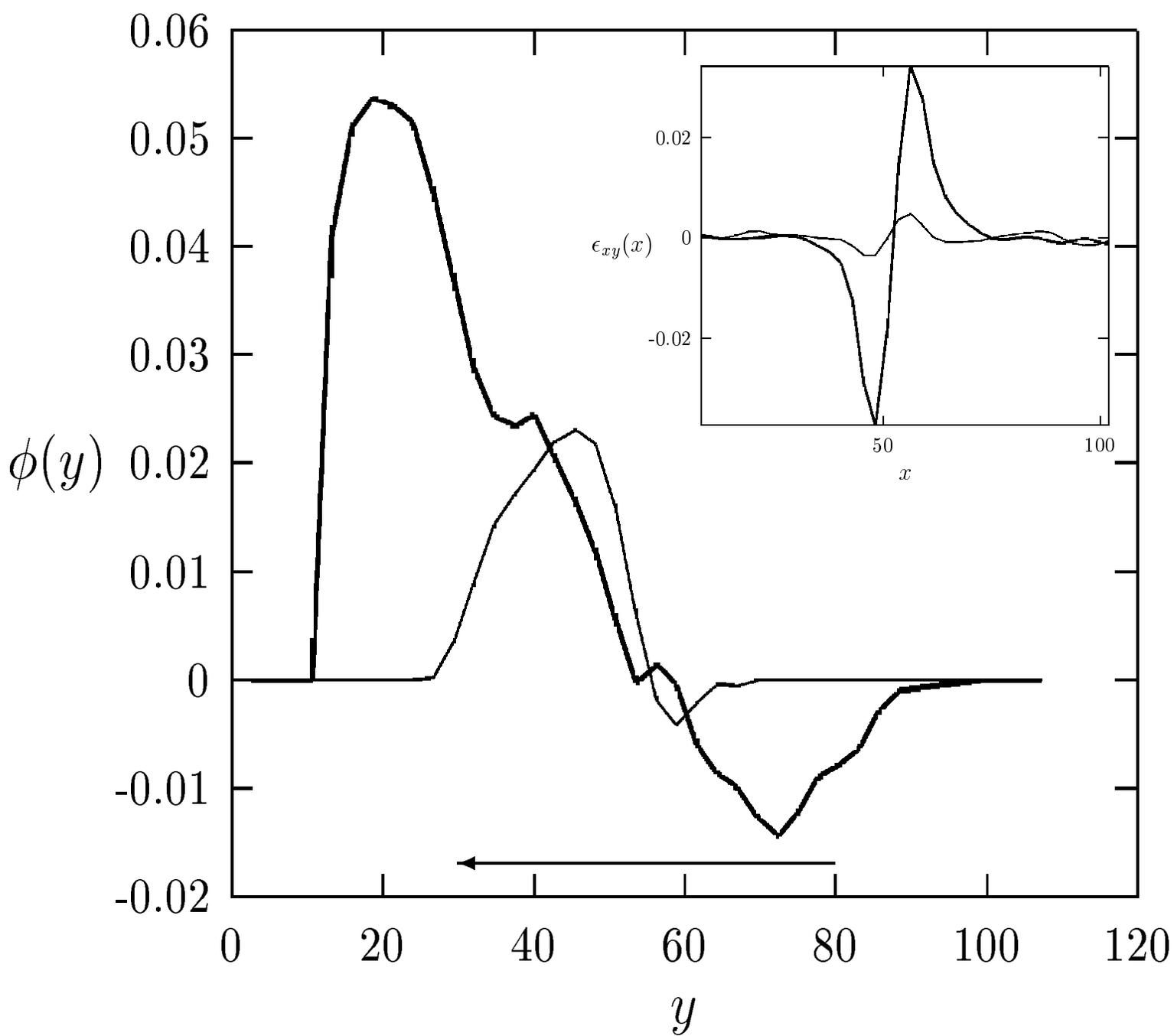
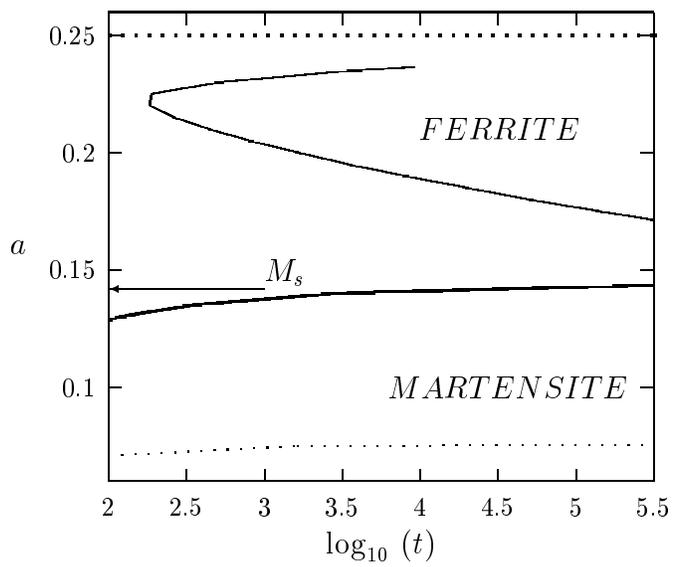
\end{document}